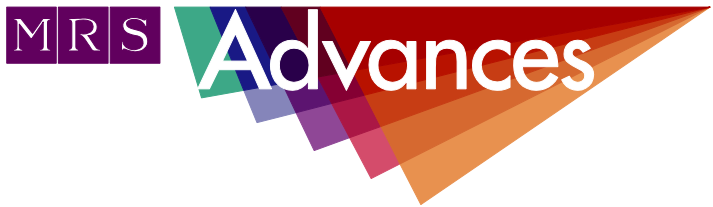

# Morphology Control in van der Waals Epitaxy of Bismuth Telluride Topological Insulators

Celso I. Fornari[1], Eduardo Abramof[1], Paulo H. O. Rappl[1], Stefan W. Kycia[2], and Sérgio L. Morelhão[3]

[1]National Institute for Space Research, São José dos Campos, SP, Brazil

[2]Department of Physics, University of Guelph, Guelph, ON, Canada

[3]Institute of Physics, University of São Paulo, São Paulo 05508-090, Brazil

## ABSTRACT

Bismuth telluride have regained significant attention as a prototype of topological insulator. Thin films of high quality have been investigated as a basic platform for novel spintronic devices. Low mobility of bismuth and high desorption coefficient of telluride compose a scenario where growth parameters have drastic effects on structural and electronic properties of the films. Recently [J. Phys. Chem. C 2019, 123, 24818−24825], a detailed investigation has been performed on the dynamics of defects in epitaxial films of this material, revealing the impact of film/substrate lattice misfit on the films' lateral coherence. Very small lattice misfit (<0.05%) are expected to have no influence on quality of epitaxial system with atomic layers weakly bonded to each other by van der Waals forces, contrarily to what was observed. In this work, we investigate the correlation between lattice misfit and size and morphology of the film crystalline domains. Three-dimensional reciprocal-space maps of film Bragg reflections obtained with synchrotron X-rays are used to visualize the spatial conformation of the crystallographic domains through film thickness, while atomic force microscopy images provide direct information of the domains morphology at the film surface.

## INTRODUCTION

Epitaxial films of bismuth telluride have been extensively investigated in the last few years [1-6]. One of the most challenging aspect in controlling structural and electronic properties of this material is the competition between desorption of tellurium and interlayer atomic mobility during growth. In molecular beam epitaxy (MBE), the key control parameters are the substrate temperature $T$ and the ratio $\Phi$ between beam equivalent pressures of Te and $Bi_2Te_3$ sources [6,7]. At temperatures below a certain value, for a given ratio of equivalent pressures, there is formation of Te-rich phases together with the $Bi_2Te_3$ phase. Immediately above this temperature to avoid Te-rich phases, the $Bi_2Te_3$ films are full of point defects and twinned domains. By further increasing the temperature, desorption of tellurium kicks in and $Bi_2Te_{3-\delta}$ films with deficit $\delta$ of tellurium are obtained, but it can be prevented to some extent by increasing the pressure of Te. Film lattice parameter, mechanical and electrical properties are impacted by composition, and the mean lattice coherence length of the films have shown direct correlation with the in-plane lattice mismatch [1]. However, in term of device processing it is also important to understand how the morphologies of crystallographic domains are related to growth conditions.

Purely morphological probes such as atomic force microscopy (AFM), besides being restricted to surface structures, they are blind to crystallographic orientation and lattice perfection of the structures. On the other hand, X-ray diffraction is ideal for crystallographic analysis, but with little susceptibility to morphology of domains unless they have sizes in nanometer length scales. In this work, we optimize X-ray diffraction tools to visualize the three-dimensional conformation of the crystallographic domains as a function of growth parameters in samples with well know degrees of twinning, composition, and lattice mismatch. For such thin films (~160 nm thick), X-ray probe analyzes equally almost the whole thickness, while AFM images are used to visualize the film morphology of the top most atomic layers. By comparing three-dimensional reciprocal space maps and AFM images, a better understanding is achieved on how to interpret the data obtained by these methods when analyzing epitaxial films of bismuth telluride.

## EXPERIMENTAL

Bismuth telluride epitaxial films were grown on freshly cleaved (111) $BaF_2$ substrates using a Riber 32P molecular beam epitaxial system. Effusion cells charged with a nominal stoichiometric $Bi_2Te_3$ solid source and two extra Te sources were used here. The solid sources are produced at our laboratory using commercially available Bi (99.999%) and Te (99.9999%) elements. The beam equivalent pressure (BEP) of the effusion cells was monitored by a Bayer-Alpert ion gauge. The extra tellurium supply $\Phi = \sum BEP_{Te} / \sum BEP_{Bi_2Te_3}$ is determined as the ratio between the Te and $Bi_2Te_3$ BEP [8]. In this case, $\Phi = 0$ indicates that no extra tellurium is provided during the growth, that is films are grown by using only the $Bi_2Te_3$ cell. The background pressure of the growing chamber never exceeded $10^{-9}$ Torr during growth. The film surface is monitored in situ during growth by reflection high-energy electron diffraction (RHEED) equipment, using a 35 keV electron cannon. All films were grown for 2h at a constant rate of 0.22 Å/s, resulting in thicknesses of 160±10 nm, as verified either by X-ray reflectometry or cross-sectional scanning electron microscopy (Tescan MIRA3 model) [6]. Other film properties that have been characterized elsewhere in samples grown under the same conditions are summarized in Table I, including the lateral lattice coherence length determined from the full width at half maximum at grazing incidence diffraction geometry [1].

Table I: Sample labels, substrate temperature (*T*), and ratio Φ of beam equivalent pressures between Te and $Bi_2Te_3$ sources. In-plane mismatch (Δ*a*/*a*), lateral lattice coherence length (*L*), film composition, and degree of twinning ($d_{tw}$) obtained elsewhere [1] for samples grown on the same conditions are also shown.

| Sample | *T* (°C) | Φ | Δ*a*/*a* (%) | *L* (nm) | Film composition | $d_{tw}$ (%) |
|---|---|---|---|---|---|---|
| S1 | 250 | 1 | -0.068(9) | 60(6) | $Bi_2Te_3$ | 19.8(0.3) |
| S2 | 270 | 1 | -0.011(4) | 158(10) | $Bi_2Te_{2.74}$ | 35.5(0.4) |
| S3 | 290 | 1 | +0.001(3) | 165(14) | $Bi_2Te_{2.58}$ | 2.7(0.1) |
| S4 | 270 | 2 | -0.056(5) | 81(7) | $Bi_2Te_3$ | 44.3(0.2) |

X-ray diffraction data were acquired at the XRD2 beamline of the Brazilian Synchrotron Light Laboratory (LNLS) [9-11]. The photon energy was tuned to 8 keV (λ = 1.54009 Å exactly) by using a double-crystal (111) Si monochromator; logitudinal coherence length of ~0.6 µm. The beam was vertically focused into the sample with a bent Rh-coated mirror placed before the monochromator; transversal coherence lengths of ~0.8 µm (vertical) and ~0.4 µm (horizontal). Spot size at the sample position: 0.6 mm (vertical) and 2 mm (axial). The sample was mounted on a Huber 4+2-circle diffractometer and the diffracted beam was analyzed by a Pilatus 100k area detector (172 µm pixel) mounted in the 2θ arm of the diffractometer. The distance between the sample and the detector was set to $D = 910$ mm. To minimize absorption and scattering of the diffracted beam into the air, an evacuated tube was installed between sample and detector.

Three-dimensional reciprocal space maps (RSMs), that is the 3D reconstruction of diffracted intensities around reciprocal lattice nodes, were carried out as follow [12,13], see details in Figure 1. In the reference frame of the laboratory where $\hat{x}$ is downstream along the incident X-ray beam, $\hat{y}$ rest on the horizontal plane, and $\hat{z}$ lays in the vertical scattering plane, the incident wavevector is simply $\boldsymbol{k} = (2\pi/\lambda)\hat{x}$ while the diffracted wavevector towards the central pixel of the detector area is $\boldsymbol{k}'_d = (2\pi/\lambda)[\cos\theta_d \cos\varphi_d\,\hat{x} + \cos\theta_d \sin\varphi_d\,\hat{y} + \sin\theta_d\,\hat{z}] = (2\pi/\lambda)\hat{s}_d$. $\theta_d$ and $\varphi_d$ stand for the angles of elevation and azimuth of the detector arm, respectively. For a detector area perfectly perpendicular to $\hat{s}_d$, as shown in Figure 1b, and pixel arrays well aligned along vertical and horizontal directions, a position vector $\boldsymbol{r}_d(m,n) = p[(m-m_0)\hat{x}_d + (n-n_0)\hat{y}_d]$ can be ascribed to each pixel regarding the position of the central pixel of array indexes $m_0 n_0$. The pixel size is $p = 0.172$ mm for the used detector, $\hat{x}_d = -\sin\theta_d \cos\varphi_d\,\hat{x} - \sin\theta_d \sin\varphi_d\,\hat{y} + \cos\theta_d\,\hat{z}$, and $\hat{y}_d = -\sin\varphi_d\,\hat{x} + \cos\varphi_d\,\hat{y}$. In the lab frame, the absolute position of each pixel is therefore given as $\boldsymbol{R} = D\hat{s}_d + \boldsymbol{r}_d(m,n)$ where *D* is the sample detector distance. The diffracted X-ray intensity at a pixel of indexes *mn* and wavevector $\boldsymbol{k}' = (2\pi/\lambda)\,\boldsymbol{R}/|\boldsymbol{R}|$ has diffraction vector $\boldsymbol{Q} = \boldsymbol{k}' - \boldsymbol{k}$. The intensity distribution is obtained after projecting $\boldsymbol{Q}$ in a convenient frame of the sample's reciprocal space, for instance $\hat{x}_s = \cos\theta\,\hat{x} + \sin\theta\,\hat{z}$, $\hat{y}_s = \hat{y}$, and $\hat{z}_s = -\sin\theta\,\hat{x} + \cos\theta\,\hat{z}$ in which $\boldsymbol{Q}_0 = Q_0\hat{z}_s$ (Figure 1a). For each step of the incidence angle *θ*, that is when rocking the sample around $\hat{y}_s$ in increments of Δθ (= 0.01° in this work), the intensity distribution in the vicinity of vector $\boldsymbol{Q}_0$ is therefore given as a function of $Q_x = \boldsymbol{Q}\cdot\hat{x}_s$, $Q_y = \boldsymbol{Q}\cdot\hat{y}_s$, and $Q_z = \boldsymbol{Q}\cdot\hat{z}_s$. The computer codes used to generate the 3D RSMs are based on the MatLab routine exrlp3dview.m, available for free download (back matter) at the publisher's website of ref. [13].

## RESULTS AND DISCUSSIONS

Three-dimensional distribution of X-ray diffracted intensities around the symmetric 00 15 reflection of the films are presented in Figure 2 along with AFM images of the films exposed surfaces. As the film hexagonal unit cell along the $c$ axis is composed of 15 atomic layers, the 00 15 reflection is at $Q_z = 2\pi/\langle d \rangle$ where the mean interlayer distance $\langle d \rangle = 2.035 - 0.025\delta$ Å varies with the deficit $\delta$ of Te [6,7].

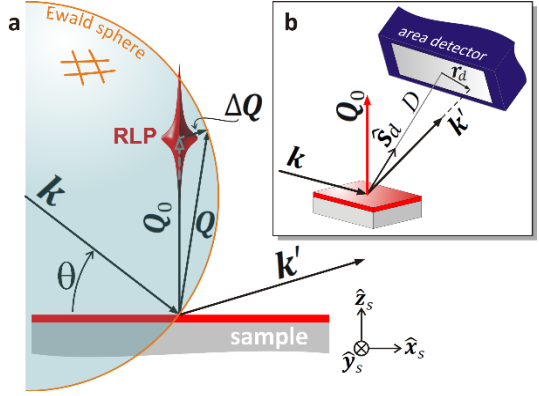

Figure 1. (a) Ewald sphere construction for describing diffracted X-ray intensities around a reciprocal lattice point (RLP) given by vector $Q_0$. For each angle of incidence $\theta$, all vectors $Q = k - k'$ ending on the surface of the Ewald sphere are diffracting, although with different intensities. The 3D intensity distribution around the RLP are given as a function of $\Delta Q = Q - Q_0$ projections in the sample's reference frame of base $\hat{x}_s$, $\hat{y}_s$, and $\hat{z}_s$. (b) Wavevector $k' = (2\pi/\lambda)[D\hat{s}_d + r_d]/|D\hat{s}_d + r_d|$ of the diffracted intensity impinging a pixel at position $r_d$ in the detector area.

In sample S1, Figure 2a, because the film is formed by domains of small sizes from the bottom to the top, the reciprocal shape and orientation of the domains are clearly visible. The AFM image shows triangular shapes with dimension smaller than 100 nm, perfectly compatible with the lateral lattice coherence length $L = 60$ nm as determined in a similar film. The low-intensity pattern in Figure 2a have six tips as if formed by two interposed pyramids, a larger one that is upside down and a smaller one rotated by 60° around the film growth direction as clearly seen in top view (insets). This low-intensity pattern can be related to surface features. But, as the intensity scale up, only one pyramidal pattern remains, see the innermost-red isointensity surface in Figure 2a. It corresponds to the main crystallographic domains in the film. Then, twinned domains that are estimated to be about 20% in this film can be responsible for the rotated portion of the low-intensity pyramidal pattern. In this case, and based on the spreading of diffracted intensities in the $Q_x Q_y$ plane, the minimum size of twinned domains seems to be 30% larger than the minimum size of the non-rotated domains.

In the AFM image of sample S2, Figure 2b, the length scale of surface structures is also of the order of 100 nm but no well-defined shapes or sharp edges have been identified. The intensity pattern in the RSM is narrower, suggesting that diffracting domains are larger than in sample S1. Moreover, the lowest isointensity surface (outermost-blue surface) is smoother and without pronounced tips as in the previous sample. Although a significant degree of twinning is expected in this film, nearly 36%, the only evidence of twinned domains seems to be a hexagonal-like intensity pattern, instead of triangular ones, as noticed in top view from the $Q_z$ direction (insets of Figure 2b). The well-defined hexagonal intensity pattern observed in this film can be understood as the smallest domains of both types (normal and twinned) having very similar sizes and shapes, both producing upside down pyramidal-like intensity distributions. The only main difference between them is the 60° rotation.

Large triangular structures with sharp edges are observed in the AFM image of sample S3, Figure 2c, as often reported in epitaxial films of bismuth telluride [6,7,14,15]. Most structures with lateral sizes close to 100 nm take place at the top of the uppermost layers where each step of 1 nm in height corresponds to one Te:Bi:Te:Bi:Te quintuple layer (QL) [7,16]. Crystallographic domains of dimensions larger that the X-ray coherence lengths, of about $0.6 \pm 0.2$ μm, produce intensity patterns with no information of shape although the low-intensity (outermost-blue surface) of the RSM in Figure 2c presents some features related to domain shapes. In top view (insets), the pattern is like a triangle with bulging sides. Because the estimated degree of twinning in this sample S3 is practically one order of magnitude smaller than in the other samples, the low-intensity pattern observed is probably caused by the contribution of the small lateral sizes of the QLs at the tip of the pyramids, as well as by all the sharp triangle corners whose edges are within the x-ray coherence length. The bulging sides of the intensity pattern may arise due to twinned domains with shapes not well defined as those seen in the AFM image of this sample S3 (Figure 2c).

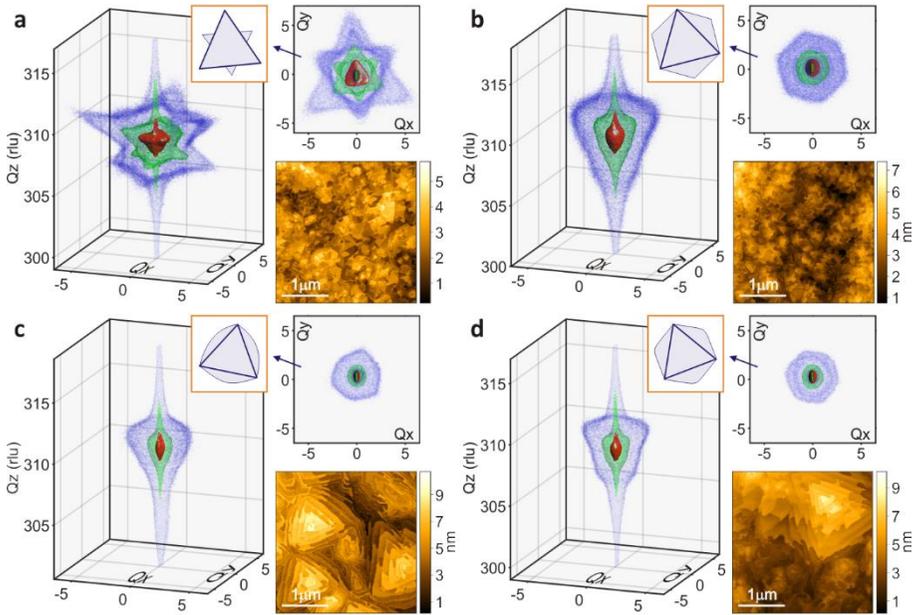

Figure 2. Reciprocal space maps (3D and top-view) and AFM images of bismuth telluride films. (a-d) Films of same thicknesses (160±10 nm) and different growth temperature T and ratio Φ of extra tellurium (Table I): (a) T = 250ºC and Φ = 1; (b) T = 270ºC and Φ = 1; (c) T = 290ºC and Φ = 1; and (d) T = 270ºC and Φ = 2. Film Bragg reflection 00 15, specular diffraction geometry, and X-rays of 8 keV. Intensity in log scale, and $Q_x$, $Q_y$, and $Q_z$ in reciprocal lattice units rlu = 0.01 Å$^{-1}$. Isointensity surfaces stand for 1,2% (outermost-blue), 4,2% (intermediate-green), and 14,4% (innermost-red) of the maximum intensities. Main features of the lowest intensity patterns (outmost-blue isointensity surface) are depicted within arrow pointed areas. Contributions from the smallest domains are indicated by triangles.

In Figure 2d, sample S4, the AFM image shows a mix of large triangles along with irregular structures. The low-intensity pattern (outermost-blue surface) of the RSM is similar to both of those obtained from samples S2 and S3, but with slight differences. The pattern is somehow intermediary, where the bulging sides of the S3 pattern in top view are more pronounced towards a hexagonal pattern, but not as well defined as in the S2 pattern. It is consistent with normal domains having better defined triangular shapes than the twinned domains that are accounting for nearly 50% of the diffracting structures in this film. Moreover, from the perspective of the narrower spreading of diffracted intensities in the $Q_xQ_y$ plane, the smallest domains in sample S4 are larger than in sample S2.

Films grown on the same conditions of the films in samples S2 and S3 have very small lattice mismatch as a consequence of variation in composition, and nearly the same lateral lattice coherence length (Table I). Despite these similarities, their surface morphologies are completely different, see AFM images in Figures 2b and 2c. It suggests that the growth temperature is one of the key parameters in defining the surface structures in films with hundreds of QLs. The perfect lattice matching in sample S3 may also has contributed to the significant reduction in the degree of twinning and improving the formation of large and uniform structures at the film surface. Withdrawing straightforward correlations between growth parameters and film properties have been aimed by several researches in the literature [4,7,8,17,18]. It is a complex problem demanding more elaborated structural probes than a few currently available. AFM as the main tool for morphological analysis of surface structures, electron microscopy for high resolution analysis of local atomic structures, and X-ray diffraction as a general tool for analyzing composition and crystallographic orientation of the domains. The 3D RSM analysis proposed here offers the possibility of combining, in a single non-destructive tool, crystallographic and morphological analysis throughout the whole film thickness. No sample conditioning is required, and data acquisition per sample takes no more than a few minutes, which is important in studies of large ensemble of samples prepared under a variety of conditions.

Another observation is that the low-intensity patterns reported here around a symmetric Bragg reflection of the films have shown a poor correlation with the expected lattice coherence length $L$ that, in principle, defines the portion of high intensity of the diffraction patterns around the reciprocal lattice nodes. For instance, at 50% of the intensity maxima the node width (fwhm) in the $Q_x$ direction is expected to follow $\Delta Q_x = 2\pi/L =$ 0.01 Å$^{-1}$, 0.004 Å$^{-1}$, 0.004 Å$^{-1}$, and 0.008 Å$^{-1}$ for samples S1 to S4, respectively; the fwhm in the $Q_y$ direction is caused by the horizontal beam size at the detector area. In qualitative disagreement with these values, there is the result from sample S2 that displays wider intensity distributions in the $Q_x$ direction than the ones of samples S3 and S4. Compare for instance the $Q_x$ width of the innermost-red isointensity surfaces of these samples. This disagreement may arise from the fact that the actual $L$ value for sample S2 can be different from the expected one, or that the RSM of the chosen symmetric Bragg reflection has different susceptibility to the $L$ value determined when using asymmetric reflections [1]. In either case, the low-intensity patterns were shown to be dominated by shape and orientation of the smallest sized domains present in the film, contrarily to the fwhm that is determined by size distribution weighted functions [19,20].

## CONCLUSIONS

A detailed description on how three-dimensional reciprocal space mapping of symmetry Bragg reflections can be used as controlling tool of shape, size, and orientation of crystallographic domains in epitaxial films of bismuth telluride were presented in this work. It is a tool available in high flux synchrotron facilities and suitable for studying epitaxial systems based on weak van der Waals forces where the absence of strong atomic interlayer forces leads to films with 2D structures highly dependent of the growth parameters. The main advantage of this tool regarding surface structure probes such as atomic force microscopy is that the RSM is not limited to the morphologic aspect of the surface, although there are correlations as demonstrated here. On the other hand, its probing capability lose efficiency in films with laterally large structures, depending on the coherence length of the X-ray beam. Advanced X-ray sources of enhanced properties of coherence may be able to burst the morphology probing efficiency of 3D RSM.

## ACKNOWLEDGMENTS

The authors acknowledge the financial support from Brazilian agencies CAPES (grant no. 88881.119076/2016-01), FAPESP (grant nos. 2016/22366-5 and 2018/00303-7), and CNPq (grant nos. 302134/2014-0, 309867/2017-7, 305764/2018-7, and 452287/2019-7), as well as from the NSERC of Canada.